\documentstyle[preprint,aps,epsf,floats]{revtex}

\begin{document}

\draft

\tighten

\preprint{\vbox{\hfill hep-ph/0001090 \\
          \vbox{\hfill January 2000} \\
          \vbox{\hfill Revised March 2000} \\
          \vskip0.8in 
         }}

\title{Thermal Abundances of Heavy Particles}

\author{Mark Srednicki\footnote{E--mail: \tt mark@vulcan.physics.ucsb.edu}}

\address{Department of Physics, University of California,
         Santa Barbara, CA 93106 }

\maketitle

\begin{abstract}
\normalsize{
Matsumoto and Yoshimura [hep-ph/9910393] have argued that there are loop 
corrections to the number density of heavy particles (in thermal equilibrium 
with a gas of light particles) that are not Boltzmann suppressed by a factor of
$e^{-M/T}$ at temperatures $T$ well below the mass $M$ of the heavy particle.
We argue, however, that their definition of the number density does not 
correspond to a quantity that could be measured in a realistic experiment.
We consider a model where the heavy particles carry a conserved U(1) charge, 
and the light particles do not.  The fluctuations of the net charge in a 
given volume then provide a measure of the total number of heavy particles 
in that same volume.  We show that these charge fluctuations are Boltzmann 
suppressed (to all orders in perturbation theory).  Therefore, we argue, 
the number density of heavy particles is also Boltzmann suppressed.
}
\end{abstract}

\vskip0.5in
\centerline{Physical Review D, in press}


In a series of papers, Matsumoto and Yoshimura 
(hereafter MY) \cite{my1,my2,my3}
have challenged the conventional wisdom concerning the number density
of a gas of heavy particles in thermal equilibrium with a gas of 
light (or massless) particles.  They consider a model of a heavy 
spin-zero boson (represented by a real scalar field $\varphi$)
interacting with a massless spin-zero boson (represented
by a real scalar filed $\chi$).  The scalar potential is 
\begin{equation}
V(\varphi,\chi)=  {\textstyle{1\over2}}M^2\varphi^2
                 +{\textstyle{1\over24}}\lambda_\varphi\varphi^4
                 +{\textstyle{1\over24}}\lambda_\chi\chi^4
                 +{\textstyle{1\over4}}\lambda \varphi^2\chi^2,
\label{v}
\end{equation}
where the couplings are all real and positive, and we take
$\lambda_\varphi \ll \lambda^2$,
$\lambda \ll \lambda_\chi$, and
$\lambda_\chi < 1$.
This hierarchy among the couplings allows the $\chi$ particles to function
as an efficient heat bath for the $\varphi$ particles.

At the level of free field theory, the equilibrium number density
of $\varphi$ particles is given by
\begin{equation}
n_{\varphi 0} = g \int {d^3p\over(2\pi)^3}\,f(E),
\label{n0}
\end{equation}
where $g=1$ counts the number of species of $\varphi$ particles,
\begin{equation}
f(E) = {1\over e^{\beta E} -1 }
\label{f}
\end{equation}
is the Bose distribution function, 
$\beta=1/T$ is the inverse temperature, and 
$E=({\bf p}^2+M^2)^{1/2}$ is the single-particle energy. 
For $T\ll M$, we have
\begin{equation}
n_{\varphi 0} = (2\pi)^{-3/2} (M T)^{3/2} e^{-\beta M};
\label{n02}
\end{equation}
the factor of $e^{-\beta M}$ means that $n_{\varphi 0}$ is
{\it Boltzmann suppressed}.  However, MY argue that there are
loop corrections to $n_{\varphi 0}$ that are not 
Boltzmann suppressed; specifically, they find \cite{my2,my3}
\begin{equation}
n_{\varphi} = n_{\varphi 0} + c\lambda^2 T^6/M^3 + \ldots
\label{n}
\end{equation}
for $T\ll M$, where $c=1/69120$, and the ellipses stand
for all higher-order corrections.  

They key issue that we wish to address (raised also in \cite{ss})
is the underlying definition of $n_\varphi$.  For $T\ll M$, MY define 
$n_\varphi$ via $n_\varphi=\rho_\varphi/M$, where
$\rho_\varphi$ is the energy density of the $\varphi$
particles.  This energy density is in turn defined 
(for all temperatures) via
\begin{equation}
\rho_\varphi = \langle{\cal H}_\varphi\rangle 
             = {\mathop{\rm Tr} {\cal H}_\varphi e^{-\beta H}
                       \over
                       \mathop{\rm Tr} e^{-\beta H} }
                      - \langle 0|{\cal H}_\varphi|0\rangle,
\label{rho}
\end{equation}
where $H$ is the total hamiltonian, and
\begin{equation}
{\cal H}_{\varphi} = {\textstyle{1\over2}}\dot\varphi^2 +
                     {\textstyle{1\over2}}(\nabla\varphi)^2 + 
                     {\textstyle{1\over2}}M^2\varphi^2 +
                     \hbox{counterterms}
\label{hphi}
\end{equation}
is the free-field part of the $\varphi$ hamiltonian,
plus counterterms (some of which involve the $\chi$ field)
that are necessary to remove infinities in this composite operator.

Eq.~(\ref{rho}) is a highly plausible definition of $\rho_\varphi$.
However, it does not correspond in any obvious way to how the
number density of $\varphi$ particles would be determined
experimentally.  Standard methods all involve
a search for individual, on-shell $\varphi$ particles.
Real-world examples of this include present-day
dark matter searches, and measurements of the
cosmic microwave background radiation.  

What is needed theoretically is a measurable attribute that 
is carried by the $\varphi$ particles only.
To create one, we modify the model slightly by making the
$\varphi$ field complex, and requiring its interactions
to conserve the corresponding U(1) charge.  We leave the
$\chi$ field real, and the $\chi$ particles neutral.
The modified scalar potential is 
\begin{equation}
V(\varphi,\chi)=  M^2\varphi^\dagger\varphi
                 +{\textstyle{1\over4}}\lambda_\varphi
                                      (\varphi^\dagger\varphi)^2
                 +{\textstyle{1\over24}}\lambda_\chi\chi^4
                 +{\textstyle{1\over2}}\lambda
                                       \varphi^\dagger\varphi\chi^2.
\label{v2}
\end{equation}
We can now study the net charge contained in a large but
finite volume $V$.  Of course, in thermal equilibrium, the
average net charge $\langle Q\rangle$ vanishes, 
but it has nonzero fluctuations $\langle Q^2\rangle$.
If we weakly gauge the U(1) symmetry with a small (and
therefore dynamically irrelevant) gauge coupling $e \ll\lambda$,
we can in principle measure these charge fluctuations without 
tracking individual $\varphi$ particles.

For $T\ll M$, it is easy to compute $\langle Q^2\rangle_0$, where 
the subscript 0 indicates that we are (for now) neglecting interactions.
The number $N_+$ of positively charged particles in a volume $V$ 
is then controlled by a Poisson distribution; this implies 
$\langle N_+^2\rangle_0 - \langle N_+\rangle_0^2 = \langle N_+\rangle_0$.
The number $N_-$ of negatively charged particles
is controlled by an independent Poisson distribution, with 
$\langle N_-^2\rangle_0 - \langle N_-\rangle_0^2 = \langle N_-\rangle_0$.
Overall charge neutrality implies 
$\langle N_-\rangle_0 = \langle N_+\rangle_0 = {1\over2}n_{\varphi 0}V$,
where $n_{\varphi 0}$ is now given by Eq.~(\ref{n0}) with $g=2$.
The net charge is $Q=N_+-N_-$, and so
\begin{eqnarray}
\langle Q^2\rangle_0 &=& \langle(N_+-N_-)^2\rangle_0
\nonumber \\
                   &=& \langle N_+^2\rangle_0 + \langle N_-^2\rangle_0
                       -2 \langle N_+\rangle_0\langle N_-\rangle_0
\nonumber \\                   
                   &=& n_{\varphi 0} V.
\label{q2}
\end{eqnarray}
We see that the charge fluctuations give us a measurement of the
total number of $\varphi$ particles in a given volume.  At higher
temperatures (but still ignoring interactions), quantum effects
modify this result to
\begin{equation}
\langle Q^2\rangle_0 = 2V \int {d^3p\over(2\pi)^3}\,f(E)[1+f(E)].
\label{q22}
\end{equation} 
The ratio $\langle Q^2\rangle_0/n_{\varphi 0}V$ depends weakly on
temperature; it rises slowly from one at $T\ll M$ to $\pi^2/6\zeta(3)=1.37$ 
at $T\gg M$.  (The result is similar for fermions, with a ratio
of one at low temperatures and $\pi^2/9\zeta(3)=0.91$ 
at high temperatures.)  
Furthermore, it seems highly unlikely that weak interactions could
significantly modify Eq.~(\ref{q2}).  If we were to have either
$\langle Q^2\rangle \ll n_\varphi V$ or
$\langle Q^2\rangle \gg n_\varphi V$, we would be forced to 
conclude that the movements of positive and negative particles
are highly correlated (in order to suppress or enhance the charge 
fluctuations in any particular volume).  This is inconsistent 
with the usual notion of a gas of particles that move freely and
independently between occasional scatterings, 
and would appear to require strong interactions.

We therefore propose to {\it define\/} the number density of
$\varphi$ particles, for $T\ll M$, via
\begin{equation}
n_\varphi = \langle Q^2\rangle/V.
\label{nphi}
\end{equation}
This definition has the advantage (not shared by the
definition used by MY) of being directly connected to
the experimentally measurable quantity $\langle Q^2\rangle$.
Adopting Eq.~(\ref{nphi}) as our definition of $n_\varphi$, 
the question becomes whether or not the loop corrections 
to $\langle Q^2\rangle$ are Boltzmann suppressed.

To compute these loop corrections,
we introduce a chemical potential $\mu$ and the partition function 
\begin{equation}
Z=\mathop{\rm Tr}e^{-\beta(H - \mu Q)}.
\label{z}
\end{equation}
We then use 
\begin{equation}
\langle Q^2\rangle = {1\over\beta^2}{\partial^2\over\partial\mu^2}\ln Z
                        \biggr|_{\mu=0}.
\label{q23}
\end{equation}
At the one-loop level (that is, for free $\varphi$ particles, and
ignoring the $\mu$-independent contributions of the $\chi$ particles), 
we have the textbook formula
\begin{equation}
\ln Z_0 = V\int{d^3p\over(2\pi)^3}\ln[1+f(E-\mu)] + (\mu\to-\mu).
\label{lnz}
\end{equation}
Using Eq.~(\ref{lnz}) in Eq.~(\ref{q23}) yields Eq.~(\ref{q22}).

\begin{figure}
\epsfxsize=6.5in
\epsfbox{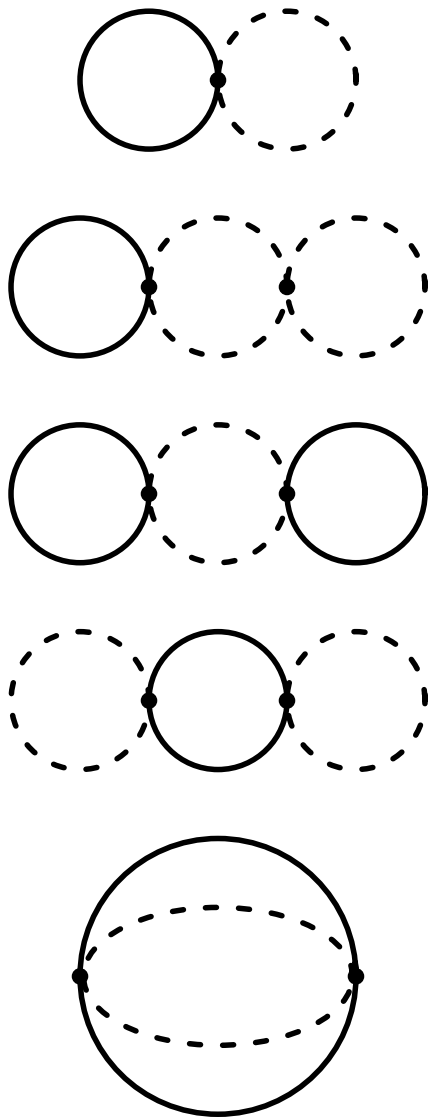}
\vspace{3mm}
\caption{Two- and three-loop contributions to $\ln Z/\beta V$ 
that include propagators for both the heavy $\varphi$ particle
(solid lines) and the massless $\chi$ particle (dashed lines).}
\label{fig1}
\end{figure}

As in \cite{my2}, loop corrections to $\ln Z$ may be computed via finite 
temperature perturbation theory (see, e.g., \cite{kap}).  
At the two- and three-loop level, the contributing diagrams 
are shown in Fig.~(1).  Vertices joining four $\varphi$ lines are 
neglected since we have assumed that $\lambda_\varphi \ll \lambda^2$.  
Furthermore we neglect diagrams with only $\chi$ lines, 
since these are independent of $\mu$. 
The $\varphi$ propagator is
\begin{equation}
\Delta(n,{\bf p})={1\over (2\pi n/\beta - i\mu)^2 + {\bf p}^2+M^2},
\label{prop}
\end{equation}
where $n$ is an integer specifying the discrete energy.
(With $M=\mu=0$, this is also the $\chi$ propagator.)  
The direction of the flow of charge is the same as the flow
of energy.  A closed $\varphi$ loop with a single vertex
contributes a factor of
\begin{equation}
T\sum_{n=-\infty}^{+\infty}\Delta(n,{\bf p})
= {1\over2E}\Bigl[1+f(E+\mu)+f(E-\mu)\Bigr],
\label{loop}
\end{equation}
where $E=({\bf p}^2+M^2)^{1/2}$.  
The ``1'' in square brackets yields a divergent, temperature independent
term that must be removed by renormalization.  The remaining two
terms are Boltzmann suppressed, since at low temperatures
\begin{equation}
f(E\pm\mu) \simeq e^{-\beta M}e^{\mp\beta\mu}e^{-\beta{\bf p}^2/2M}.
\label{bs}
\end{equation}  
This implies that the two-loop diagram in Fig.~(1),
and the first two three-loop diagrams, are all Boltzmann suppressed. 
The third three-loop diagram has a factor of $\sum_n\Delta(n,{\bf p})^2$,
and turns out to be suppressed as well.  This leaves only the last
(``basketball'') diagram.

According to the standard rules of finite temperature perturbation 
theory \cite{kap}, the contribution of the basketball diagram is
\begin{eqnarray}
{1\over \beta V}\ln Z_{\rm bb} &=& C \lambda^2 T^4
\sum_{n\hbox{{}'}\rm s}
\int{d^3p_1\over(2\pi)^3}\,
    {d^3p_2\over(2\pi)^3}\,
    {d^3k_3\over(2\pi)^3}\,
    {d^3k_4\over(2\pi)^3}\,
(2\pi)^3\delta^3({\bf p}_1+{\bf p}_2+{\bf k}_3+{\bf k_4})
\nonumber \\
&& \times\beta\delta_{n_1+n_2+n_3+n_4,0}\Delta_1\Delta_2\Delta_3\Delta_4 .
\label{bb}
\end{eqnarray}
Here ${\bf p}_1$ and ${\bf p}_2$ are the momenta on the $\varphi$
lines, and ${\bf k}_3$ and ${\bf k}_4$ are the momenta on the $\chi$
lines.  Each momentum and energy is taken to flow from the left vertex
to the right vertex.  Thus we have, in the individual propagators,
masses $M_1=M_2=M$ and $M_3=M_4=0$, and chemical potentials
$\mu_1=-\mu_2=\mu$ and $\mu_3=\mu_4=0$.  Note especially that
$\mu_2=-\mu_1$; this is because, with our convention on energy flow,
the charge flow must be opposite to the energy flow on one of the 
$\varphi$ lines, and this changes the sign of the chemical potential.
Finally,
\begin{equation}
C={1\over2!}\cdot\left(-{1\over2}\right)^2\cdot 2={1\over4}
\label{c}
\end{equation}
is a combinatoric factor.  The $1/2!$ comes from the expansion of
$\exp(-\beta H_{\rm int})$,
the $(-1/2)^2$ comes from two vertex factors arising from
the last term in the scalar potential, Eq.~(\ref{v2}), 
and the $2$ comes from the number of ways to match
up the $\chi$ lines.

Eq.~(\ref{bb}) can be evaluated by the standard procedure \cite{kap}
of writing the Kroneker delta as an integral,
\begin{equation}
\beta\delta_{n_1+\ldots+n_4,0}
=\int_0^\beta dt\;e^{2\pi i(n_1+\ldots+n_4)t/\beta},  
\label{delta}
\end{equation}
and then performing each sum via contour integration,
\begin{eqnarray}
\sum_{n=-\infty}^{+\infty}{e^{2\pi i n t/\beta} \over
                           (2\pi n/\beta - i\mu)^2 + E^2}
&=&{\beta\over 2\pi}\oint dz\,{e^{izt}\over e^{i\beta z}-1}\,
                              {1\over (z-i\mu)^2+E^2}
\nonumber \\
&=& {\beta\over 2E}\left[  e^{(\beta-t)(E+\mu)}f(E+\mu) 
                         + e^{t(E-\mu)}f(E-\mu) \right].
\label{sum}
\end{eqnarray}
Here the $z$ contour originally encloses the real axis, and then
is deformed into two small circles surrounding the poles at 
$z=i(\mu\pm E)$; this deformation is allowed for $0\le t \le \beta$.
Putting all of this together we now have
\begin{equation}
{1\over \beta V}\ln Z_{\rm bb} = C \lambda^2 
\int \widetilde{dp}_1\,\widetilde{dp}_2\,
     \widetilde{dk}_3\,\widetilde{dk}_4\,
     (2\pi)^3\delta^3({\bf p}_1+{\bf p}_2+{\bf k}_3+{\bf k_4})
     \int_0^\beta dt\;D_1 D_2 D_3 D_4,
\label{bb2}
\end{equation}
where $\widetilde{dp}_i=d^3p_i/(2\pi)^3(2E_i)$, and
\begin{equation}
D_i=\left[  e^{(\beta-t)(E_i+\mu_i)}f(E_i+\mu_i) 
                       + e^{t(E_i-\mu_i)}f(E_i-\mu_i) \right].
\label{d}
\end{equation}
The integral over $t$ is tedious but straightforward to perform;
the result is
\begin{eqnarray}
\int_0^\beta dt\;D_1 D_2 D_3 D_4
&=&
\sum_{\delta_1=0}^1 \sum_{\delta_3=0}^1 \sum_{\delta_4=0}^1
\Biggl[
{ (\delta_1+f_{1+}) (\delta_1+f_{2-}) (\delta_3+f_3) (\delta_4+f_4)
\over
\varepsilon_1 E_1 + \varepsilon_1 E_2 + \varepsilon_3 E_3 + \varepsilon_4 E_4 }
\nonumber \\ 
&& {} + 
{ (\delta_1+f_{1+}) (1-\delta_1+f_{2+}) (\delta_3+f_3) (\delta_4+f_4)
\over
\varepsilon_1 E_1 - \varepsilon_1 E_2 + \varepsilon_3 E_3 + \varepsilon_4 E_4 }
\Biggr] + (\mu \rightarrow -\mu),
\label{bb3}
\end{eqnarray}
where $\varepsilon_i=(-1)^{1-\delta_i}$, and $f_{1+}=f(E_1+\mu)$,
$f_{2-}=f(E_2-\mu)$, $f_3=f(E_3)$, etc.  We have made repeated
use of the relation
\begin{equation}
e^{\beta E}f(E) = 1 + f(E)
\label{ff}
\end{equation}
in obtaining Eq.~(\ref{bb3}).

We are interested only in those terms (if any) in Eq.~(\ref{bb3})
that are not Boltzmann suppressed.  As we see from Eq.~(\ref{bs}),
any term containing a factor of either $f_{1\pm}$ or $f_{2\pm}$ 
is Boltzmann suppressed. Thus, we can drop all such terms.  
At this point, we see immediately that the remaining terms
are all independent of $\mu$, since the only $\mu$ dependence 
in Eq.~(\ref{bb3}) is that which is contained implicitly in the factors of 
$f_{1\pm}$ and $f_{2\pm}$.  Therefore, all $\mu$-dependent contributions
(at the two- and three-loop level) to $\ln Z$ are
Boltzmann suppressed.  Eqs.~(\ref{q23}) and (\ref{nphi}) then imply
that $\langle Q^2\rangle$ and $n_\varphi$ are also Boltzmann suppressed.
This is our main result.

For completeness, and for comparison with the results of MY \cite{my2},
we continue the computation of the unsuppressed terms in $\ln Z_{\rm bb}$.
We set $f_{1\pm} = f_{2\pm} = 0$.  All numerators in Eq.~(\ref{bb3})
are then either $1$, $f_3$, $f_4$, or $f_3 f_4$.  The first of these
yields a divergent, temperature independent term that is canceled
by a renormalization of the vacuum energy. 
The second and third yield divergent, temperature dependent 
terms that are canceled by renormalization of the $\varphi$
and $\chi$ thermal self-energies (see the corresponding discussion
for QED in \cite{kap}).  Dropping these terms, we have
\begin{equation}
\int_0^\beta dt\;D_1 D_2 D_3 D_4 \to 
                                 \sum_{\varepsilon_3 = \pm1}
                                 \sum_{\varepsilon_4 = \pm1}
                                 { 2 f_3 f_4 \over
                                  E_1+E_2
                                  +\varepsilon_3 E_3+\varepsilon_4 E_4}. 
\label{bb4}
\end{equation}
Substituting this into Eq.~(\ref{bb2}), making a low temperature
expansion, and dropping a final divergent contribution to the
vacuum energy ultimately yields
\begin{equation}
{1\over \beta V}\ln Z_{\rm bb} 
   = {\pi^2 \lambda^2 T^8 \over 648000 M^4 } + O(T^{10}).
\label{bb5}
\end{equation}
The corresponding correction to the total energy density $\rho$ 
is then given by
\begin{equation}
\delta\rho
= {}-{1\over V}{\partial\over\partial\beta}\ln Z_{\rm bb}
= {7 \pi^2 \lambda^2 T^8 \over 648000 M^4 } + O(T^{10}).
\label{rhobb}
\end{equation}
This is consistent with the results of MY \cite{my2}, 
who also find that the leading correction to $\rho$ is of order $T^8/M^4$.

That there are corrections of this form to $\rho$ is not surprising.
Without interactions, we have $\rho=(\pi^2/30)T^4$ from the $\chi$ particles,
plus the Boltzmann-suppressed contribution of the $\varphi$ particles.
At temperatures $T\ll M$, we should be able to integrate the heavy 
$\varphi$ field out of the functional integral, and be left with an 
effective lagrangian for the massless $\chi$ field alone.  
This lagrangian will contain nonrenormalizable interaction terms
that are suppressed by powers of $\lambda$ and inverse powers of $M$.
These will give rise to corrections like Eq.~(\ref{rhobb}).
From this point of view, it is clear that all such corrections should be
thought of as modifications of the energy of the $\chi$ particles,
and not as unsuppressed contributions to the energy of the 
$\varphi$ particles.

We now argue that $\langle Q^2\rangle$ (and hence $n_\varphi$)
is Boltzmann suppressed to all orders in perturbation theory.
Consider an exact evaluation of the partition function 
\begin{equation}
Z=\sum_\alpha e^{-\beta(E_\alpha-\mu Q_\alpha)},
\label{z2}
\end{equation}
where the sum is over a basis of energy and charge eigenstates.
States that yield a $\mu$-dependent contribution to $Z$ must 
have $Q\ne0$.  States consisting of a single stable $\varphi$
particle are well defined exact energy eigenstates, 
with $E=({\bf p}^2+M^2)^{1/2}$ and $Q=\pm 1$;
their contribution to $Z$ is obviously Boltzmann suppressed.  
Furthermore, any other state with $Q=\pm1$ must have energy 
$E>M$; otherwise, the $\varphi$ particle would not be stable
(since it could decay into this lighter charged state). 
At tree level, energy eigenstates with two or more $\varphi$ 
particles all have $E\ge 2M$; interactions can modify this
to $E\ge 2M[1+O(\lambda^2)]$.  The $O(\lambda^2)$ term 
could in principle be negative (if there is a bound state of
two or more like-charge particles), but cannot make $E \ll 2M$
in a weakly coupled theory.
We therefore conclude that all $\mu$-dependent contributions 
to $Z$ are Boltzmann suppressed.  We have already seen this 
explicitly at the level of two and three loops.

With this in mind, we can return to the original model of MY,
Eq.~(\ref{v}).  At the level of Feynman diagrams,
the models are essentially identical, the only difference being 
the combinatoric factor associated with each diagram.
We therefore expect that $n_\varphi$ should be Boltzmann suppressed 
in this model as well.
We are still in need, however, of a general definition of
$n_\varphi$ that does not rely on the trick of Eq.~(\ref{nphi}),
but which corresponds to experimental measurements.

We conclude, in accord with \cite{ss}, that the proper definition
of the number density $n_\varphi$ of heavy particles 
(in equilibrium with a thermal bath of light particles) 
is a delicate matter.  We have argued that it is essential
that $n_\varphi$ be defined in a manner that renders it 
measurable in a realistic experiment.  
In this paper, we have considered a model in which 
the heavy particles carry a conserved U(1) charge, 
while the light particles are neutral.
In this situation, the local charge fluctuations provide an
experimentally accessible measure of 
the number density of heavy particles.
We have shown explicitly that, at low temperatures, 
this number density is Boltzmann suppressed up through
three-loop order in perturbation theory, and we have argued
that this must in fact be true to all orders. 
   
\begin{acknowledgments}

I thank Anupam Singh and Scott Thomas for discussions.
This work was supported in part
by the National Science Foundation through grant PHY--97--22022, 
and by the Institute of Geophysics and Planetary Physics through grant 920.

\end{acknowledgments}

\end{document}